\documentclass[preprint,authoryear,5p,times,twocolumn]{elsarticle}

\usepackage{txfonts}
\usepackage{graphicx}
\usepackage{longtable}
\usepackage{color}
\usepackage{multirow}
\usepackage{natbib}
\usepackage{pdfpages}
\usepackage{subfigure}

\usepackage{savesym}
\savesymbol{iint}
\savesymbol{iiint}
\savesymbol{iiiint}
\savesymbol{idotsint}
\usepackage{amsmath}
\restoresymbol{AMS}{iint}
\restoresymbol{AMS}{iiint}
\restoresymbol{AMS}{iiiint}
\restoresymbol{AMS}{idotsint}
\bibpunct{(}{)}{;}{a}{}{,}
\newcommand{\beq}{\begin{equation}}
\newcommand{\eeq}{\end{equation}}

\newcommand{\bdi}{\begin{displaymath}}
\newcommand{\edi}{\end{displaymath}}

\usepackage{lineno,hyperref}
\modulolinenumbers[5]

\journal{Astroparticle Physics}

\begin{document}

\title{Local interstellar spectra and solar modulation of cosmic ray 
electrons and positrons} 
\begin{frontmatter}
%\author{Cheng-Rui Zhu$^{1,2}$, Qiang Yuan$^{1,3,4}$, Da-Ming Wei$^{1,3}$}
\author[lab1,lab2]{Cheng-Rui Zhu}
\author[lab1,lab3,lab4]{Qiang Yuan}
\ead{yuanq@pmo.ac.cn}
\author[lab1,lab3]{Da-Ming Wei}
\ead{dmwei@pmo.ac.cn}

\address[lab1]{Key Laboratory of Dark Matter and Space Astronomy, Purple
Mountain Observatory, Chinese Academy of Sciences, Nanjing 210033, China}
\address[lab2]{University of Chinese Academy of Sciences, Beijing 100049, China}
\address[lab3]{School of Astronomy and Space Science, University of Science and
Technology of China, Hefei 230026, Anhui, China}
\address[lab4]{enter for High Energy Physics, Peking University, Beijing 100871, China}

%\email{yuanq@pmo.ac.cn (QY), dmwei@pmo.ac.cn (DMW)}

\begin{abstract}
Low energy cosmic rays are modulated by the solar activity when they
propagation in the heliosphere, leading to ambiguities in understanding
their acceleration at sources and propagation in the Milky Way.
By means of the precise measurements of the $e^-$, $e^+$, $e^-+e^+$,
and $e^+/(e^-+e^+)$ spectra by AMS-02 near the Earth, as well as the very 
low energy measurements of the $e^-+e^+$ fluxes by Voyager-1 far away from 
the Sun, we derive the local interstellar spectra (LIS) of $e^-$ and $e^+$
components individually. Our method is based on a non-parametric 
description of the LIS of $e^-$ and $e^+$ and a force-field solar
modulation model. We then obtain the evolution of the solar modulation
parameters based on the derived LIS and the monthly fluxes of cosmic ray
$e^-$ and $e^+$ measured by AMS-02. {\bf To better fit the monthly
data, additional renormalization factors for $e^-$ and $e^+$ have been
multiplied to the modulated fluxes.} We find that the inferred solar
modulation parameters of positrons are in good agreement with that of
cosmic ray nuclei, and the time evolutions of the solar modulation 
parameters of electrons and positrons differ after the reversal 
of the heliosphere magnetic field polarity, which shows clearly the
charge-sign dependent modulation effect.
\end{abstract}
  
\end{frontmatter}
%%%%%%%%%%%%%%%%%%%%%%%%%%%%%%%%%%%%%%%%%%%%%%%%%%%%%%%%%%%%%%
\section{Introduction}
Large progresses have been achieved in the direct measurements of cosmic
rays (CR) in the past decade, by space experiments including the AMS, 
Fermi-LAT, DAMPE, CALET, and NUCLEON, providing very important information 
about the origin, acceleration, and propagation of cosmic rays in the Milky 
Way \citep[e.g.,][]{2019IJMPD..2830022G,2019PrPNP.10903710K}. Nevertheless,
there is still strong degeneracy among the acceleration and propagation
effects (including those in the heliosphere), which hinders an unambiguous 
understanding of the CR problems. Very interestingly, the Cosmic Ray 
Subsystem (CRS) instrument on the Voyager-1 spacecraft launched more than 
40 years ago keeps on operation and measuring the low-energy CR fluxes even 
outside the heliosphere\footnote{Note that it is still possible that there
is tiny residual modulation effect on the Voyager-1 spectra 
\citep{2011ApJ...735..128S,2014ApJ...782...24K}. In this work we assume 
that the Voyager-1 measurement is the LIS without considering such subtlety.} 
\citep{2013Sci...341..150S,2016ApJ...831...18C}.
In addition, the PAMELA and AMS-02 experiments further reported time
variations of the CR fluxes with very high precisions 
\citep{2013ApJ...765...91A,2016PhRvL.116x1105A,2018ApJ...854L...2M,
2018PhRvL.121e1102A,2018PhRvL.121e1101A}, which are also direct relevant 
to the solar modulation. The Voyager-1 data, and/or the time series of CR 
fluxes, are very important in probing the local interstellar spectra (LIS) 
and solar modulation effect of CRs \citep[e.g.,][]{2016Ap&SS.361...48B,
2016A&A...591A..94G,2016ApJ...829....8C,2017ApJ...840..115B,
2017ApJ...849L..32T,2018ApJ...863..119Z,2018ApJ...854...94B,
2018PhRvL.121y1104T,2019ApJ...871..253C,2019PhRvD.100f3006W}. 

In \citet{2018ApJ...863..119Z} we studied the LIS of CR nuclei from He to 
O with a non-parametric spline interpolation method and the force-field
model of the solar modulation \citep{1967ApJ...149L.115G,1968ApJ...154.1011G}, 
according to AMS-02 \citep{2017PhRvL.119y1101A,2018PhRvL.120b1101A}, 
Voyager-1 \citep{2016ApJ...831...18C}, and ACE-CRIS data. The time-evolution
of the solar modulation parameters were then derived based on the monthly
ACE-CRIS fluxes of CR nuclei, which are consistent with those inferred
from the neutron monitors \citep{2011JGRA..116.2104U,2017AdSpR..60..833G}.

In this work we extend the previous study to electrons and positrons.
One of our motivations is to examine the possible differences between 
the LIS of electrons and that of nuclei, which may have important
implication in the propagation of different particle species in the
Milky Way \citep{2015PhRvD..91f3508L}. Furthermore, the differences 
of solar modulation effects among electrons, positrons, and nuclei
may help to understand the charge-sign dependent modulation effects.

The CR electron and positron spectra and flux ratios have been 
measured precisely by several space experiments, such as PAMELA 
\citep{2009Natur.458..607A,2011PhRvL.106t1101A,2013PhRvL.111h1102A}, 
Fermi-LAT \citep{2017PhRvD..95h2007A}, AMS-02 
\citep{2014PhRvL.113l1102A,2014PhRvL.113v1102A,2014PhRvL.113l1101A,PhysRevLett.122.101101,PhysRevLett.122.041102}, 
DAMPE \citep{2017Natur.552...63D}, and CALET \citep{2018PhRvL.120z1102A}. 
The Voyager-1 experiment also measure the total $e^-+e^+$ fluxes from $\sim3$ 
to $\sim40$ MeV outside the heliosphere \citep{2016ApJ...831...18C}.
Here the data obtained on top-of-atmosphere (TOA) by AMS-02 and in the
local interstellar space by Voyager-1 will be used.
Following the method of \citet{2018ApJ...863..119Z}, we adopt a
non-parametric spline interpolation method to describe the LIS of 
electrons and positrons, which are then modulated under the force-field
model and fitted to the long-term average data. Based on the LIS derived 
above and the monthly fluxes of electrons and positrons, we then derive
the time-series of the modulation parameters. We are aware that the
force-field model should be over-simplified in modeling the solar
modulation effect. However, the more physical modulation model usually 
has a considerable number of free parameters and is computationally heavy 
\citep{2016CoPhC.207..386K,2017ApJ...839...53L,2017A&A...601A..23P,
2019PhRvD.100d3007V,Kuhlen_2019}. {\bf Thus we keep the framework of the 
force-field model, but with some extension with additional remormalization 
factors. Also the electrons and positrons have different modulation 
parameters. We expect that the extended force-field approximation can 
reasonably reflect the main features of the solar modulation.}

\begin{figure*}[!htb]
\centering
\includegraphics[width=\textwidth]{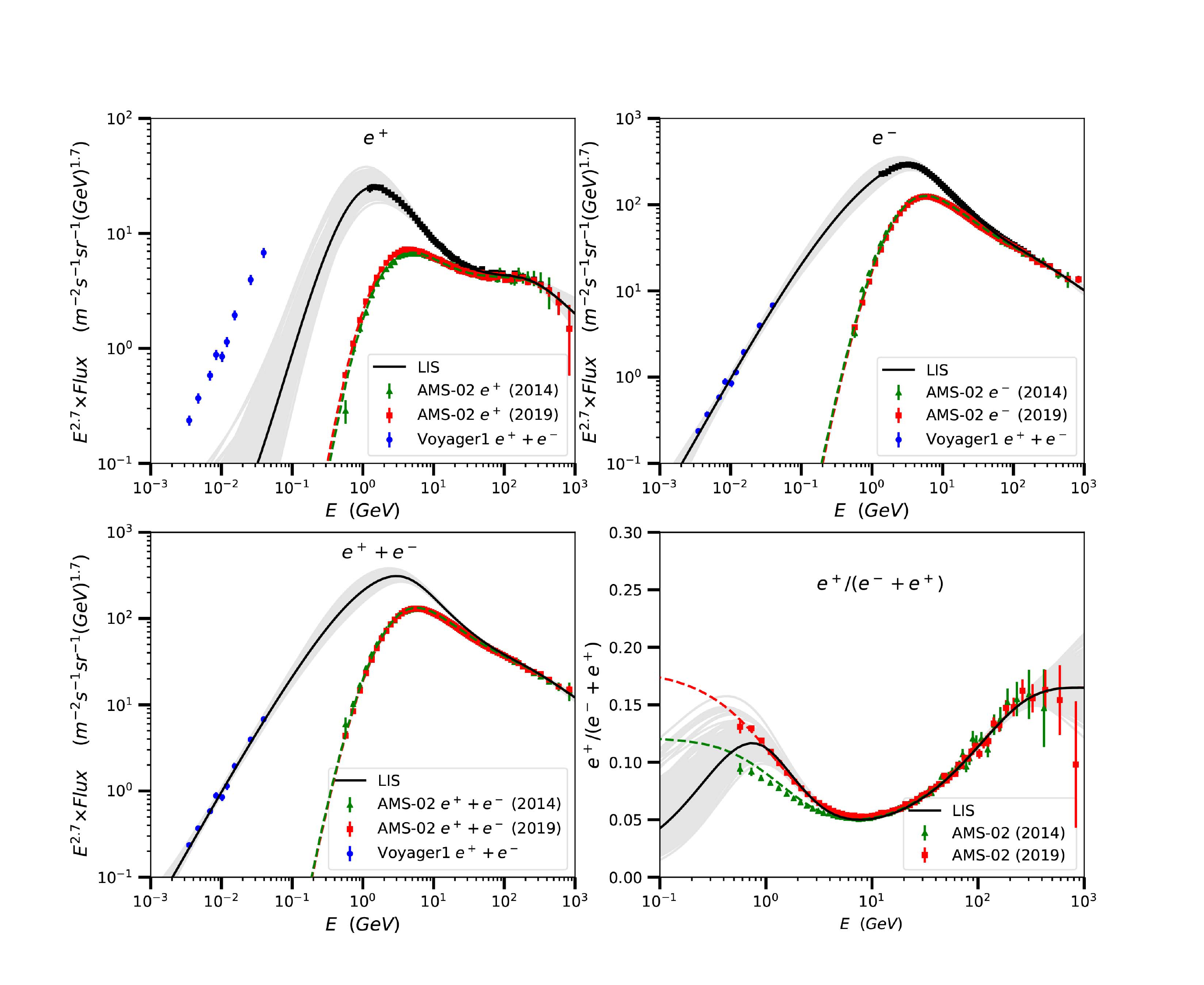}
\caption{Best-fit LIS fluxes (lines), multiplied by $E^{2.7}$, 
compared with the measurements of Voyager-1 \citep[blue points;][]
{2016ApJ...831...18C}, and the results of AMS-02 (2014) \citep[red points;][]
{2014PhRvL.113l1102A,2014PhRvL.113v1102A,2014PhRvL.113l1101A}, 
AMS-02 (2019)\citep[green points;][]{PhysRevLett.122.101101,
PhysRevLett.122.041102}. The green and red dashed linses are the best-fit
TOA spectra for AMS-02 (2014) and AMS-02 (2019) data, respectively.
The de-modulated results of the AMS-02 (2019) data are denoted by the 
black points, based on the fitted $\phi_{\pm}$ values. The solid lines
are the best-fit LIS, with the gray bands being the 68\% coverage of 
the fitting results.}
\label{fig:flux} 
\end{figure*}

%%%%%%%%%%%%%%%%%%%%%%%%%%%%%%%%%%%%%%%%%%%%%%%%%%%%%%%%%%%%%%
\section{Methodology}

Usually the CR spectra are parameterized with power-law or broken power-law
function \citep{1998ApJ...493..694M,2018ApJ...854...94B}. However, the
actual CR spectrum, either the accelerated one or the detected one, may
be more complicated. More and more new features of the CR spectra have
been revealed by recent observations \citep[e.g.,][]{2017Natur.552...63D,
2019arXiv190912860A,2010ApJ...714L..89A,2018JETPL.108....5A}.
To minimize the impact of the assumed function form of the energy spectra
of CRs, following \citet{2016A&A...591A..94G} and \citep{2018ApJ...863..119Z},
we adopt a cubic spline interpolation method to describe the wide-band
LIS of both electrons and positrons. Note that in this current work the
propagation of electrons and positrons in the Milky Way is not discussed.
The cubic spline interpolation is a way to get a smoothly connecting 
piecewise function passing through a set of energy points. We work in the 
$\log(E)-\log(J)$ space, where $E$ is the energy of electrons or positrons in 
unit of GeV and $J$ is the flux in unit of GeV~m$^{-2}$~s$^{-1}$~sr$^{-1}$. 
The selected positions of knots $\boldsymbol{x}=\log(E)$ are:
\begin{equation}
    \begin{split}
        &\{x_1,x_2,x_3,x_4,x_5,x_6,x_7,x_8\} \\
        =&\{-2.5,-1.4,0.0,0.6,1.2,1.8,2.4,3.0\}.
    \end{split}
\end{equation}
In the low energy range the knots are sparse because the data points in 
such energy ranges are very limited. The corresponding fluxes 
$y_{+,i}=\log(J_{+,i})$ and $y_{-,i}=\log(J_{-,i})$ are free
parameters to be fitted.

Since most of the observations are carried out near the Earth, they just
give the modulated TOA spectra. As we have mentioned before, we use the 
force-field solar modulation model \citep{1967ApJ...149L.115G,
1968ApJ...154.1011G} to link the LIS with the TOA spectra as
\begin{equation}\label{force_filed}
J^{\rm TOA}(E)=J^{\rm LIS}(E+\Phi)\times\frac{E(E+2m_e)}
{(E+\Phi)(E+\Phi+2m_e)}, 
\end{equation}
where $E$ is the kinetic energy of the particle, $\Phi=\phi\cdot e$ with 
$\phi$ being the solar modulation potential, $m_e=0.511$ MeV is the 
electron mass, and $J$ is the differential flux of electrons or positrons. 
Note that here the modulation parameters for electrons and positrons,
$\phi_{-}$ and $\phi_{+}$, are assumed to be independent and fitted 
simultaneously.

The $\chi^2$ statistics is defined as 
\begin{eqnarray}
\chi^2=\sum_{i=1}^{m}\frac{{\left[J_{\pm}(E_i;\boldsymbol{y}_{\pm},
\phi_{\pm}{\rm ~or~}0)-J_i\right]}^2}{{\sigma_i}^2},
\end{eqnarray}
where $J_{\pm}(E_i;\boldsymbol{y}_{\pm},\phi_{\pm})$ is the expected TOA/LIS
fluxes of $e^+$, $e^-$, $e^++e^-$ or the ratios $e^+/(e^++e^-)$, $J_i$ and 
$\sigma_i$ are the measured data and error for the $i$th data bin.

We use the Markov Chain Monte Carlo (MCMC) method \citep{2002PhRvD..66j3511L} 
to fit the parameters. The MCMC is based on the Bayesian framework which can 
minimize the $\chi^2$ function, and give the posterior distributions of the
high-dimensional parameter space with a high efficiency.
The likelihood function of the model parameters is
\begin{equation}
{\mathcal L}(\boldsymbol{\theta})\propto \exp\left(-\frac{\chi^2}{2}\right),
\end{equation}
The posterior probability of model parameters is then
\begin{equation}
p(\boldsymbol{\theta}|{\rm data}) \propto {\mathcal L}(\boldsymbol{\theta}) 
p(\boldsymbol{\theta}),
\end{equation}
where $p(\boldsymbol{\theta})$ is the prior probability of parameters
$\boldsymbol{\theta}$. Here we assume flat priors of all the parameters.

We adopt the Metropolis-Hastings algorithm which generates Markov chains 
as follows. For a set of parameters $\boldsymbol{\theta_i}$ and its
successor $\boldsymbol{\theta_{i+1}}$, we calculate an accept probability
$P_{\rm acc}=\min[{p(\boldsymbol{\theta_{i+1}}|{\rm data})/ 
p(\boldsymbol{\theta_i}|{\rm data}),~1]}$.
If $\boldsymbol{\theta_{i+1}}$ is accepted, then repeat the precedure from
$\boldsymbol{\theta_{i+1}}$. Otherwise, we go back to $\boldsymbol{\theta_i}$.
The procedure is continued until a convergence criterion is satisfied.

The data used in the fit include the TOA measurements of the $e^-$, 
$e^+$, $e^-+e^+$ fluxes and $e^+/(e^++e^-)$ ratios by AMS-02 in 2014
\citep{2014PhRvL.113v1102A,2014PhRvL.113l1102A,2014PhRvL.113l1101A}, 
{\bf and 2019 \citep{PhysRevLett.122.101101,PhysRevLett.122.041102}}, 
and the LIS of $e^-+e^+$ measured by Voyager-1 \citep{2016ApJ...831...18C}.
The LIS of both $e^+$ and $e^-$ are assumed to monotonically decrease
with energies, and the LIS of $e^+$ is further assumed to be smaller
than that of $e^-$. The latter requirement is based on the fact that
there are primary $e^-$ accelerated at the sources. The fit determines
the LIS of $e^+$ and $e^-$, and the average solar modulation potentials 
for the time from May, 2011 to November, 2013, {\bf and from May, 2011 
to November, 2017} during which the measurements of $e^-$, $e^+$, 
$e^-+e^+$ fluxes and $e^+/(e^++e^-)$ ratios by AMS-02 were made. 
After deriving the LIS through the above fits, the time-dependent 
measurements of the $e^-$ and $e^+$ fluxes for every Bartels rotation 
period ($\sim27$~days) are then used to derive the time-variation of 
the solar modulation parameters $\phi_{\pm}$.

%%%%%%%%%%%%%%%%%%%%%%%%%%%%%%%%%%%%%%%%%%%%%%%%%%%%%%%%%%%%%%%%%%%
\section{Results}

\subsection{The LIS of $e^-$ and $e^+$}

In Fig.~\ref{fig:flux} we show the fitting spectra of $e^+$ (upper left), 
$e^-$ (upper right), $e^++e^-$ (lower left) and the positron fraction 
(lower right), compared with the AMS-02 (2014)\citep{2014PhRvL.113l1102A,
2014PhRvL.113v1102A,2014PhRvL.113l1101A}, AMS-02 (2019) 
\citep{PhysRevLett.122.101101,PhysRevLett.122.041102}, and Voyager-1 
\citep{2016ApJ...831...18C} data. The gray bands show the 68\% uncertainty
ranges of the fits. Note that the Voyager-1 data are the $e^++e^-$ fluxes, 
which are also shown in the top panels for comparison. Since there are 
lack of measurements of $e^+$ and $e^-$ fluxes at low energies 
(for $E\lesssim0.5$~GeV), there are relatively large uncertainties of 
the fitting results in such an energy range. The degeneracy between the
solar modulation parameters and the low energy fluxes further enlarge
the uncertainties. Therefore we observe relatively wide bands of the
$e^+$ and $e^-$ LIS at low energies. The sum of the $e^+$ and $e^-$ is
constrained by the Voyager-1 data, and is less uncertain.
 
%The probability density functions of some selected parameters are shown 
%in Fig.~\ref{fig:pdf_corr}. They are the first, second and fifth knots 
%for electrons, the second knot for positron, and $\phi$ for both. This 
%figure shows low correlation of $\phi$ with high energy and too low 
%energy knots, because the high energy particles  suffer less effects 
%of solar modulation, while the measurements of  TOA data have low energy 
%threshold. The knots at high energy ($\geq 1 GeV$) have small uncertainties, 
%while the low energy have larger uncertainties as we have only $e^++e^-$ 
%data from Voyager-1 in this energy range. 

\begin{figure}[!htb]
\centering
\includegraphics[width=0.48\textwidth]{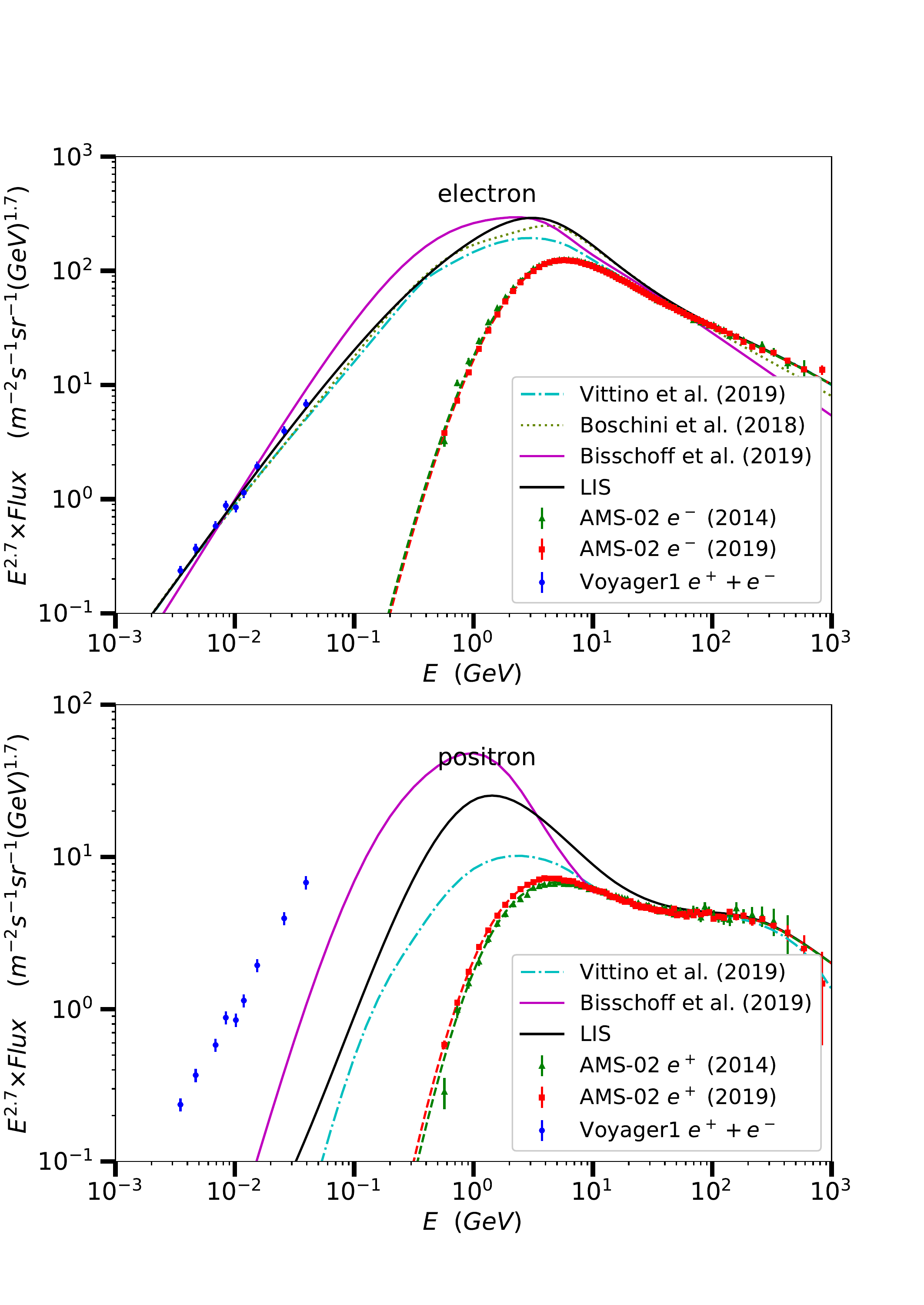}
\caption{Comparison of our derived LIS with previous works 
\citep{2019PhRvD.100d3007V,Bisschoff_2019,2018ApJ...854...94B}.}
\label{fig:compare} 
\end{figure}

In Fig.~\ref{fig:compare} we compare our best-fitting LIS of $e^-$ 
(top panel) and $e^+$ (bottom panel) with previous works with somehow
different methods and assumptions 
\citep{2019PhRvD.100d3007V,Bisschoff_2019,2018ApJ...854...94B}.
For the $e^-$ spectrum, our result is very close to that of 
\citet{2019PhRvD.100d3007V} and \citet{2018ApJ...854...94B} at low
and high energies. The main differences appear at medium energies
(from $0.1$~GeV to $10$~GeV), which show the uncertainties of the
solar modulation modelings. The result of \citet{Bisschoff_2019}
is higher than the others at lower energies (from $0.01$~GeV to 
$1$~GeV), with a harder spectrum, due to the use of a different data 
sample from Voyager-1 \cite{2013Sci...341..150S,2016ApJ...831...18C}.
The LIS of $e^+$ obtained in \citet{2019PhRvD.100d3007V,Bisschoff_2019} 
shows relatively large differences from our best-fit result but is 
still consistent with our relatively wide uncertainty band 
(see Fig.~\ref{fig:flux}).

\begin{table}[ht!]
\caption{\bf Posterior mean values and $68\%$ credible level uncertainties
of the solar modulation potentials and $\chi^2$ values of various species.}
\centering
\begin{tabular}{cccccc}
    \hline\hline
    Species & $\phi$ (GV) & $\chi^2$ \\
    \hline
    $e^-$ (2014)     & $0.762 \pm 0.039$ & 31.4 \\
    $e^+$ (2014)     & $0.754 \pm 0.039$ & 45.7 \\
    $e^++e^-$ (2014) & $ - $ & 23.3  \\
    $e^+/(e^++e^-)$ (2014) &$ - $& 124.3  \\
    $e^-$ (2019)     & $0.792 \pm 0.038$ & 26.8 \\
    $e^+$ (2019)     & $0.693 \pm 0.039$ & 40.5\\
    $e^++e^-$ (2019) & $ - $ &  22.15 \\
    $e^+/(e^++e^-)$ (2019) &$ - $& 83.1  \\    
    $e^++e^-$ (Voyager) &$ - $& 8.95  \\   

    \hline
\end{tabular}
\label{tab:chi2}
\end{table}

\begin{figure}[ht!]
\centering
\includegraphics[width=0.48\textwidth]{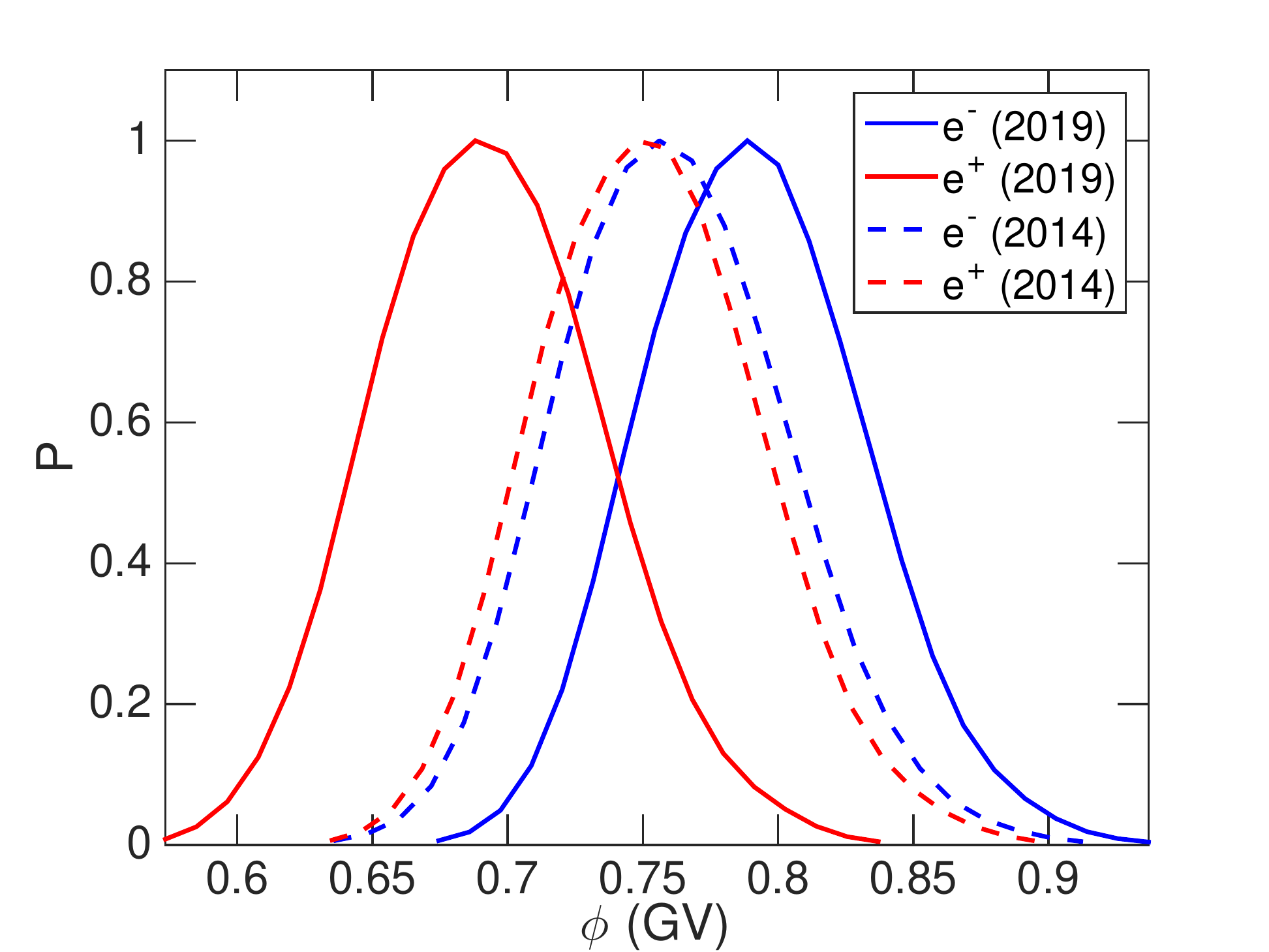}
\caption{The probability density distributions of $\phi_{-}$ (blue)
and $\phi_{+}$ (red).}
\label{fig:pdf}
\end{figure} 
    
{\bf The fitting results of the solar modulation potentials $\phi_-$ and
$\phi_+$, and the $\chi^2$ values are given in Table \ref{tab:chi2}. 
We find that the $\chi^2$ values for positron fluxes and positron 
fractions are relatively large. As we will do below, if we multiply
two renormalization factors $c_\pm$ on the 2014 data, we find
$c_-=0.997\pm0.004$ and $c_+=1.038\pm0.005$, and the fits will be 
improved significantly. This result shows that the simple force-field
modulation model may be not enough to describe the solar modulations
at different solar conditions.}

The normalized probability distributions of $\phi$ are shown in 
Fig.~\ref{fig:pdf}. The results show that the electrons were modulated 
more severely than positrons during the period from 2013 to 2017. 
From 2011 to 2014, the electrons and positrons were modulated
similarly with each other.
Based on the derived modulation potentials, we de-modulate the AMS-02 
(2019) electron and positron fluxes from the TOA to the LIS, as shown 
by the balck points in Fig.~\ref{fig:flux}. Note that the errors of
$\phi_{\pm}$ are included in the total errors of the de-modulated
fluxes via the error propagation method. The de-modulated AMS-02 data 
are provided in Tables \ref{tab:e} and \ref{tab:p} in the Appendix.

%++++++++++++++++++++++++++++++++++++++++++++++++++++++++++
%++++++++++++++++++++++++++++++++++++++++++++++++++++++++++
\subsection{Time-dependent TOA fluxes of $e^-$ and $e^+$}
 
\begin{figure*}[ht!]
\centering
\includegraphics[width=\textwidth]{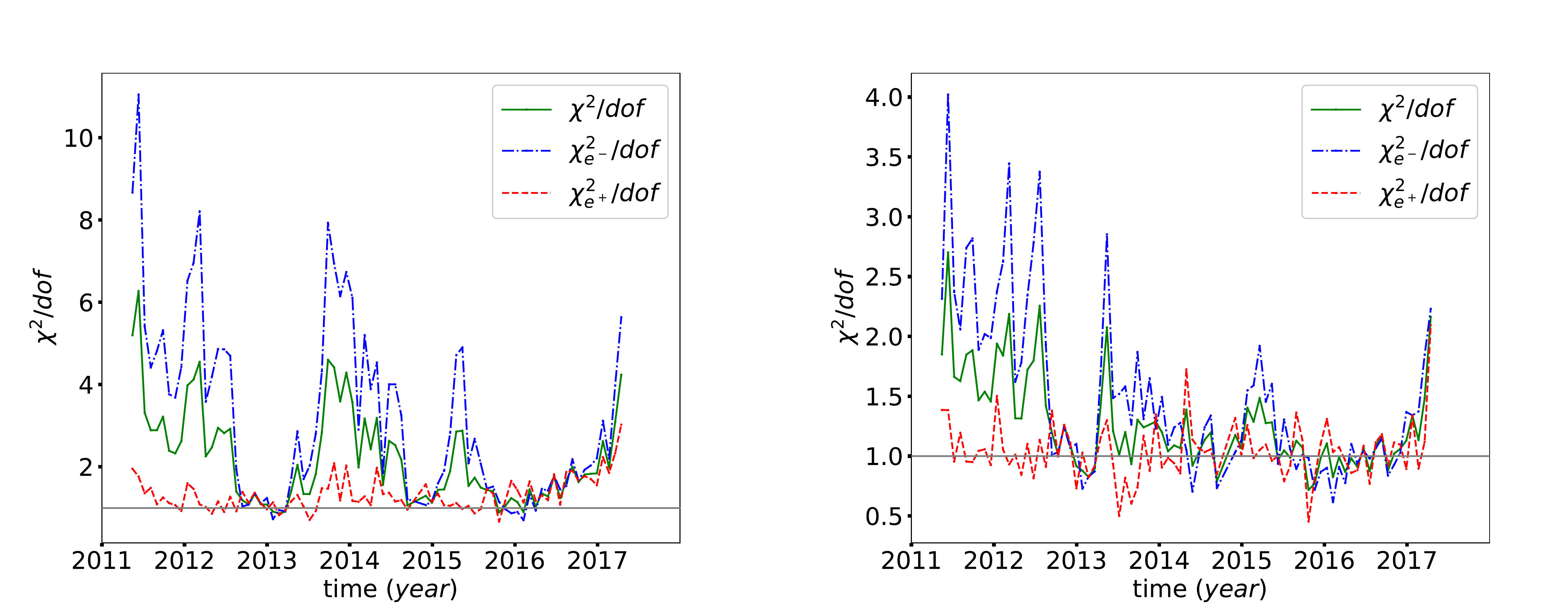}  
\caption{The $\chi^2$/dof values obtained through fitting to the 
time-dependent spectra of AMS-02. The left (right) panel is for the 
fit without (with) renormalization factors.}
\label{fig:chi2_time}
\end{figure*}

Given the LIS fluxes, we can then investigate the time-evolutions of 
the TOA fluxes, and compare them with the long-term AMS-02 measurements  
\citep{2018PhRvL.121e1102A}. However, we find that the direct fit ({\bf
with the MCMC method}) to the monthly data with the derived LIS plus a 
force-field modulation model can not always give a good fit. 
The minimum $\chi^2$ values 
diveded by the number of degree-of-freedom (dof) are shown in 
the left panel of Fig.~\ref{fig:chi2_time} with the green line.  We can see that these
fits give typically too large reduced $\chi^2$ values. 
%\textcolor{red}{As shown in \cite{Cuoco_2019}, the proton fluxes 
%from Bartels rotation 2460 and Bartels rotation 2476, cross at 4 GV. 
%This also happen in the monthly fluxes of electron and positron as 
%discussion in \cite{Kuhlen_2019}.}
This is perhaps due to the complicated perturbations of the interplanetary 
space by solar activity whose effect can not be easily accounted for by 
the simple force-field modulation model. To improve the fits of the 
time-dependent fluxes, more complicated solar modulation model and more 
free parameters are needed \citep{2019PhRvD.100f3006W,Kuhlen_2019,
2019PhRvD.100d3007V}.  
 
Empirically we extend the force-field approximation through multiplying 
two renormalization factors, $c_{\pm}$, on the LIS as
\begin{equation}
J^{\rm TOA}(E)=c_{\pm} \times J^{\rm LIS}(E+\Phi)\times\frac{E(E+2m_e)}
{(E+\Phi)(E+\Phi+2m_e)}. 
\end{equation}
The fits can be improved significantly, as shown by the green line in 
Fig.~\ref{fig:chi2_time}. 
{\bf We also show the reduced $\chi^2$ for $e^-$ and $e^+$ separately 
in Fig.~\ref{fig:chi2_time}. Generally we see that the goodness-of-fits 
for positrons are better than that for electrons.}

\begin{figure}[ht!]
\centering
\includegraphics[width=0.48\textwidth]{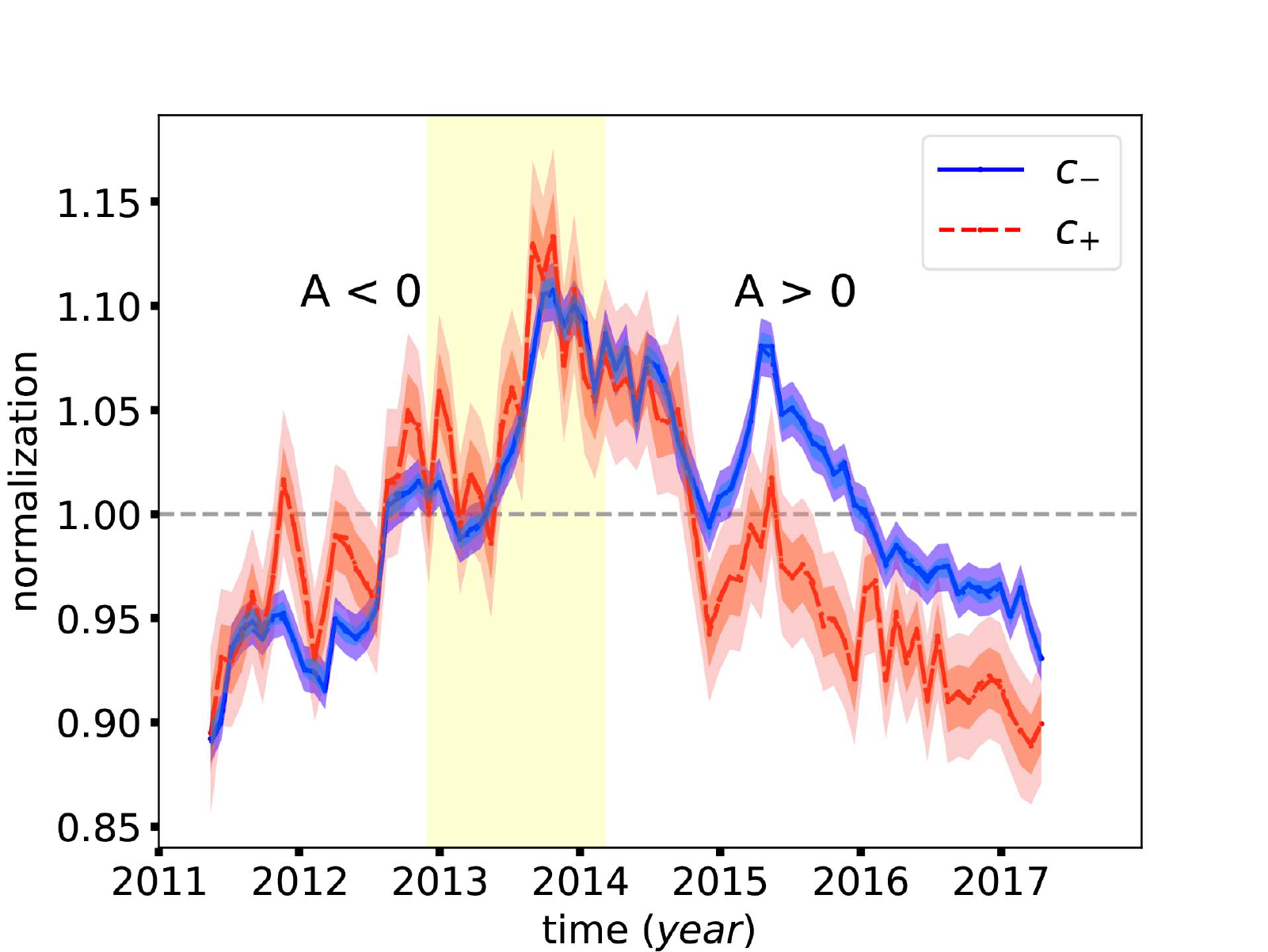} 
\caption{The renormalization factors $c_{-}$ (blue) and $c_{+}$ (red) 
for different time, the dark and light color bands stand for $1\sigma$ and $2\sigma$ credible intervals. The polarity of the heliospheric magnetic field 
is denoted by $A<0$ and $A>0$, and the yellow band stands for the 
reversal period within which the polarity is uncertain \citep{2015ApJ...798..114S}.} 
\label{fig:norm_time}
\end{figure}

\begin{figure*}[ht!]
\centering
\includegraphics[scale=0.6]{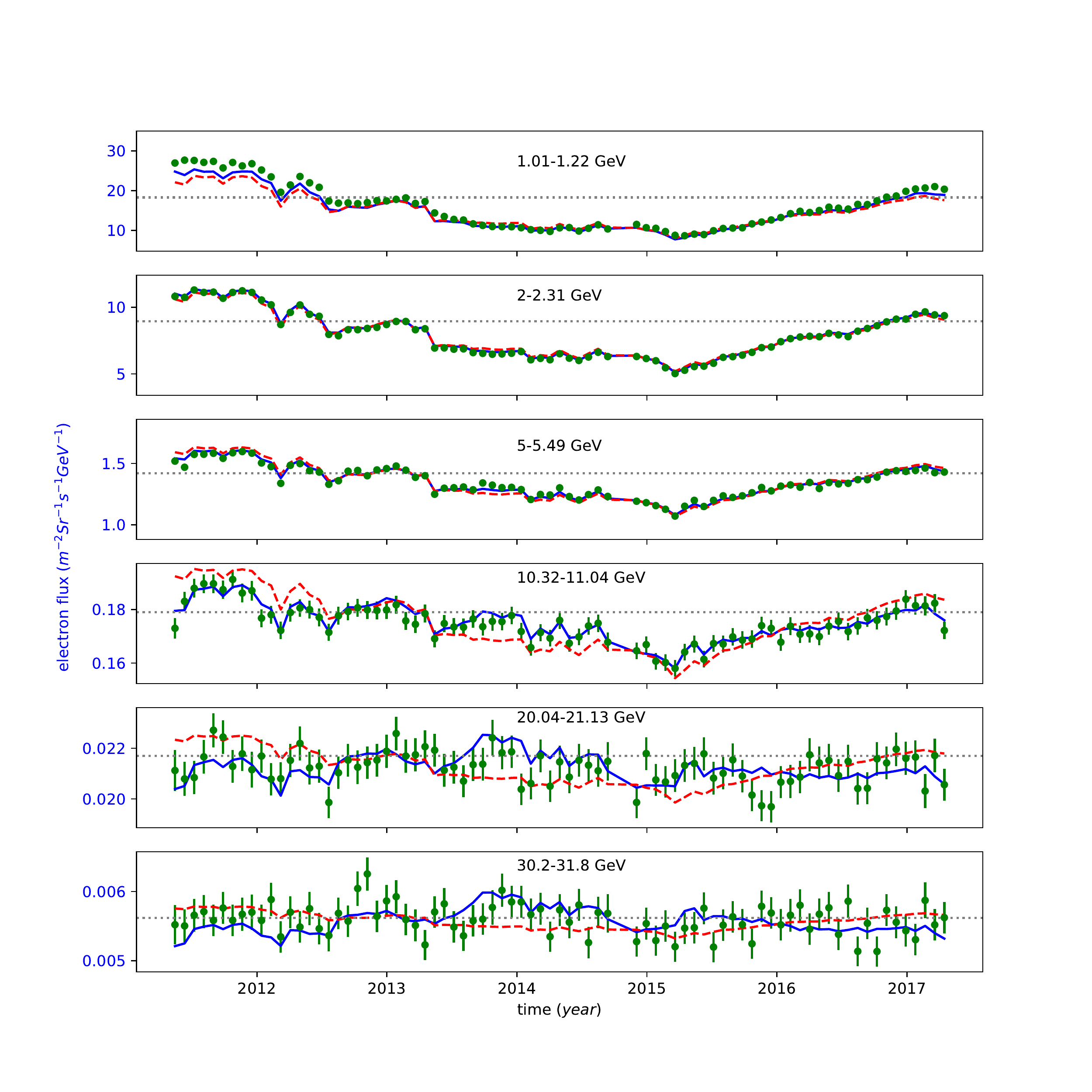}
\vspace{-1cm}
\caption{The time variations of $e^-$ fluxes for six energy bins, compared
with the AMS-02 data \citep{2018PhRvL.121e1102A}. The dot lines stand for 
the average fluxes of AMS-02 from May, 2011 to November, 2013. The blue 
solid lines are our fitting results with renormalization factors, and the 
red dash lines are those without renormalizations.}
\label{fig:flux_et}
\end{figure*}

\begin{figure*}[ht!]
\centering
\includegraphics[scale=0.6]{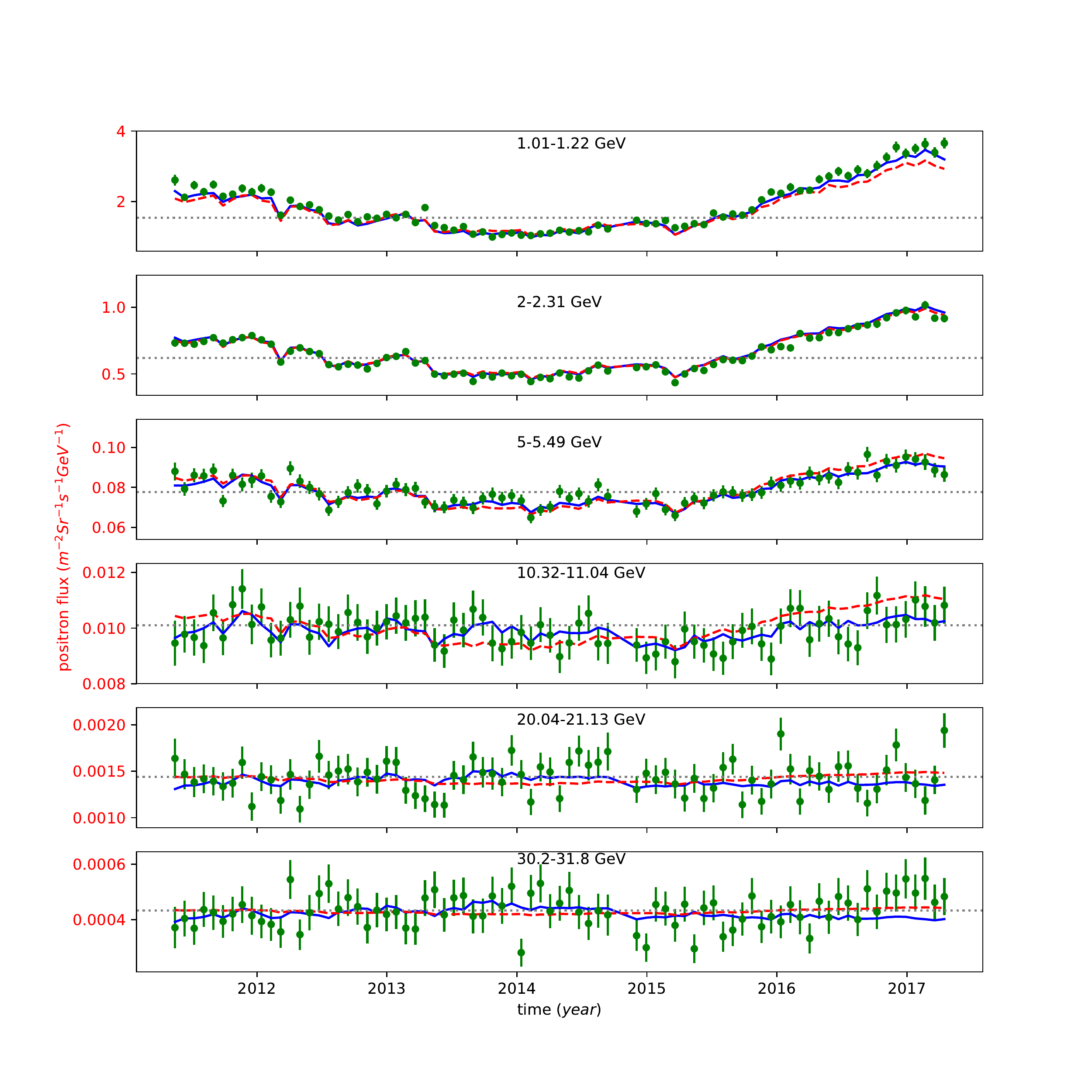}
\vspace{-1cm}
\caption{Same as Fig.~\ref{fig:flux_et}, but for positrons.}  
\label{fig:flux_pt}
\end{figure*}

\begin{figure*}[htb!]
    \centering
    \includegraphics[width=0.48\textwidth]{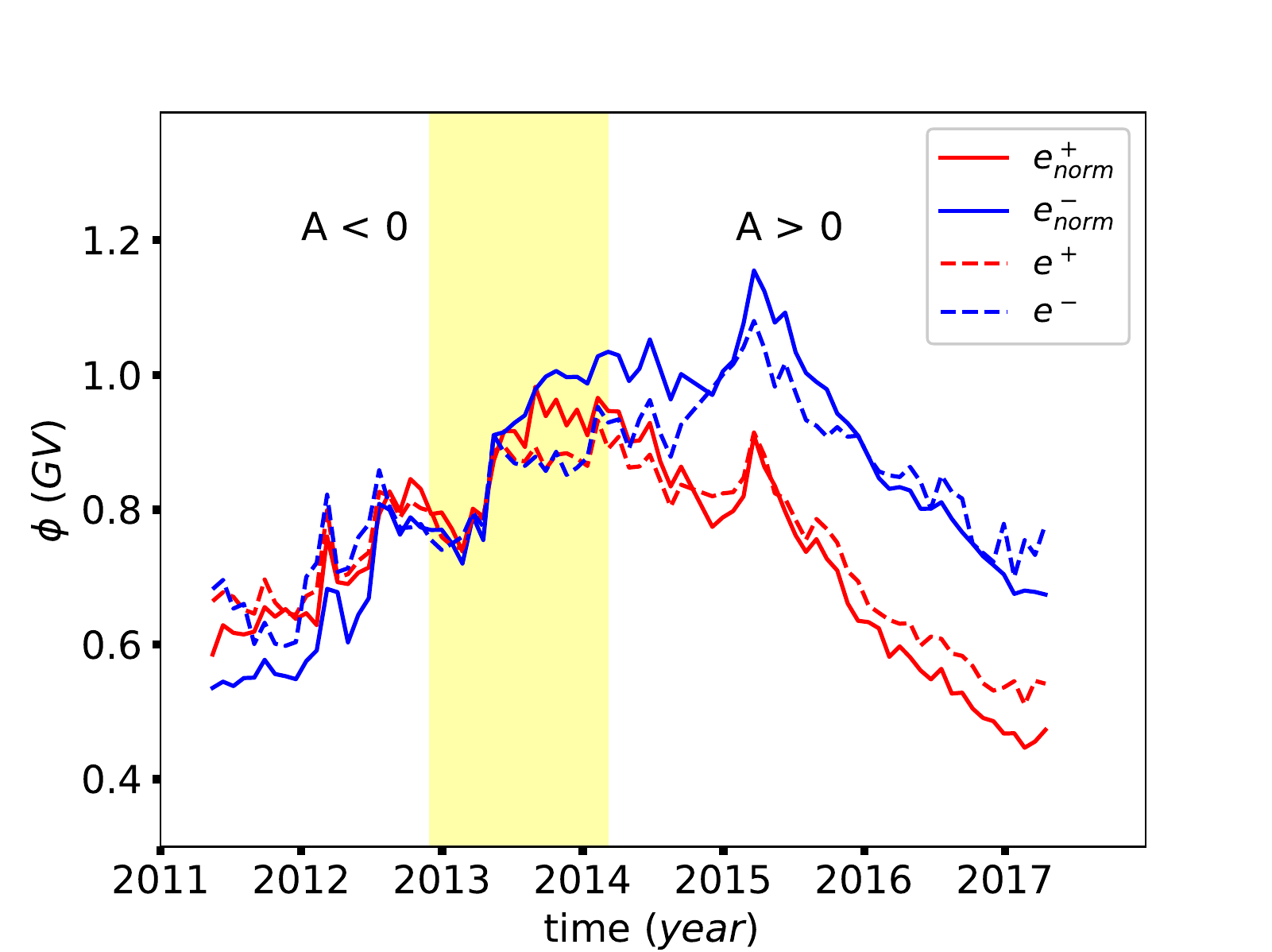} 
    \includegraphics[width=0.48\textwidth]{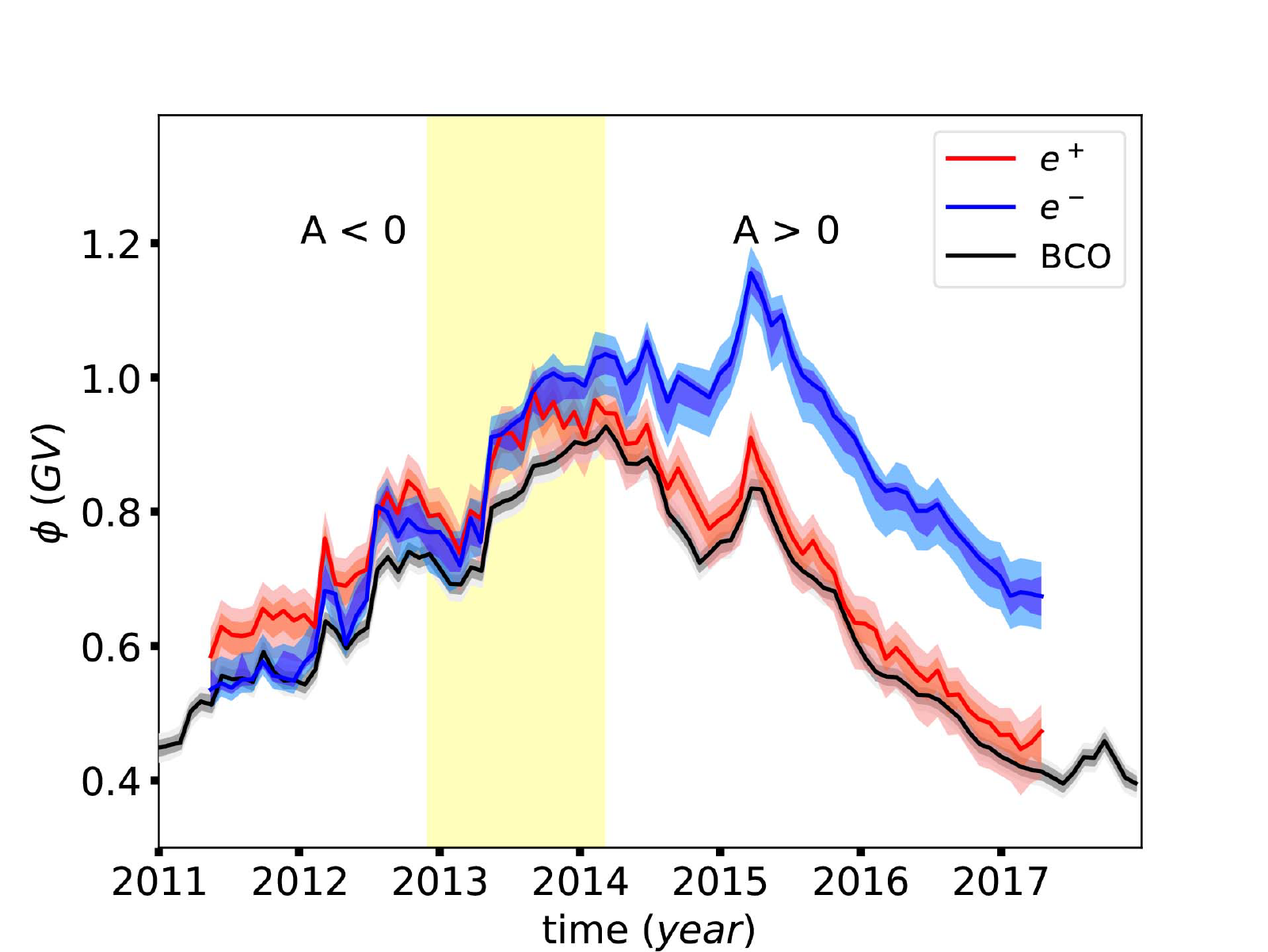} 
    \caption{Left: Time series  $\phi_{-}$ (blue) and $\phi_{+}$ (red)  
    with (solid lines) / without (dashed lines) renormalization factors 
    via fitting to the AMS-02 monthly data from 2011 to 2017.
    Right: Time series and the associated $1\sigma$ and $2\sigma$ 
    uncertainty bands of $\phi_{-}$ (blue) and $\phi_{+}$ (red) via fitting 
    to the AMS-02 monthly data from 2011 to 2017, compared with the results derived 
    from the ACE data of nuclei \citep[black;][]{2018ApJ...863..119Z}.}
    \label{fig:phi_time}
\end{figure*}

The values of $c_{\pm}$ are given in Fig.~\ref{fig:norm_time}. 
The renormalization factors show a general correlation with the 
solar activity. We expect that 
this is due to the mismatch between the energy-dependence of the 
force-field modulation model and the measurements for different time.
To see these results in more details, we plot in Fig.~\ref{fig:flux_et} 
and Fig.~\ref{fig:flux_pt} the time variations of the $e^-$ and $e^+$ 
fluxes in 6 energy bins, respectively. Taking the results before 2013 
as illustration, we can see that without including the renormalization 
factors, the model (red dashed line) predicts lower fluxes at low 
energies than the measurements and higher at high energies. In other 
words, the model spectrum is harder than the data. 
{\bf The renomalization factor $c_{\pm}<1$, together with a smaller
modulation potential $\phi_{\pm}$ compared with the cases of $c_{\pm}=1$ 
(as can be seen in Fig.~\ref{fig:phi_time}), can solve this discrepancy 
satisfactorily. Smaller $\phi_{\pm}$ gives higher low-energy fluxes,
while $c_{\pm}$ suppresses high-energy fluxes when the solar modulation
is weak. Therefore the model spectrum (blue solid lines) becomes softer
and better match the data.} Things become opposite at solar maximums, 
when $c_{\pm}>1$ is required. Furthermore, we note that differences of
the renormalization factors of $e^-$ and $e^+$ appear after the 
heliospheric magnetic field reversal. This gives a charge-sign
dependence of the solar modulation effect as expected. 

{\bf The correlation between $c_\pm$ and $\phi_\pm$ may be understood
as the drift effect of CRs in the heliosphere. As shown in
\cite{1979ApJ...234..384J} and \cite{2011ApJ...735...83S}, the presence
of drift in the Parker equation tends to give a softer TOA spectrum.
In the force-field approximation, a softer TOA spectrum means a smaller
$\phi_{\pm}$. To match with the data with relatively high energies
(above a few GeV), a renormalization factor $c_{\pm}<1$ is required.
This case applies for the periods from 2011 to 2013, and from 2015
to 2017. For the period from 2015 to 2017, the drift effect is weak,
and the modulated spectrum is thus harder. Therefore we need larger
$\phi_{\pm}$ and $c_{\pm}>1$.}

%{\bf The correlation between $c_\pm$ and $\phi_\pm$ may be understood
%as the drift effect of CRs in the heliosphere. As shown in 
%\cite{1979ApJ...234..384J} and \cite{2011ApJ...735...83S}, the presence 
%of drift in the Parker equation tends to reduce the solar modulation
%and result in a softer TOA spectrum. In the force-field approximation,
%a softer TOA spectrum means a smaller $\phi$, and hence the modulated
%spectrum is higher. In this case we need a renormalization factor 
%$c_{\pm}<1$ to match with the data.} 

\subsection{The time variation of $\phi$ }

In this sub-section, we derive the time series of the solar modulation 
potentials, according to the fits to the monthly AMS-02 data discussed
above, with the renormalization factors. To properly take into account
the uncertainties of the LIS, we adopt a Bayesian approach with the
posterior probability of $\phi_{\pm}$ being given by
\begin{equation}
p(\phi_{\pm}|{\rm data})\propto\int {\mathcal L}(\phi_{\pm},\boldsymbol{y},
c_{\pm}))\,p(\boldsymbol{y})\,p(c_{\pm})\,{\rm d}\boldsymbol{y}\,{\rm d}c_{\pm},
\end{equation}
where ${\mathcal L}$ is the likelihood of model parameters 
($\phi_{\pm},\boldsymbol{y}, c_{\pm}$), $p(\boldsymbol{y})$ is the prior 
probability distribution of $\boldsymbol{y}$ which is obtained in the 
fit in Sec.~3.1, and $p(c_{\pm})$ is the prior of $c_{\pm}$ which is
assumed to be a flat distribution within $[0.6,1.4]$.

{\bf In the left panel of Fig.~\ref{fig:phi_time}, we compare the 
fitting results of $\phi_{\pm}$ for the cases without (dashed) and 
with (solid) renormalization factors. The overall behaviors of $\phi_\pm$ 
are similar for both cases. However, small changes of the $\phi_\pm$ 
parameters appear when including the renormalization factors. Specifically,
when $c_\pm<1$ ($c_\pm>1$), $\phi_\pm$ become smaller (larger) than that
without renormalization factors.}

The right panel of Fig.~\ref{fig:phi_time} shows the time series of 
$\phi_{\pm}$ from 2011 to 2017, compared with the results derived from 
fitting to the B, C, and O nuclei data from ACE \citep{2018ApJ...863..119Z}. 
It is very interesting to find that the profile of $\phi_+$ results 
derived in this work are in good agreement with $\phi_{\rm BCO}$. 
The modulation potentials for negative charge particles, $\phi_-$, 
show systematical differences from that of positive charge particles, 
especially for $A>0$ regime. Specifically, in the $A>0$ regime, positive 
charged particles are less significantly modulated than negative charged 
particles. {\bf This could be understood as a drift effect of particles 
with different charge sign. For the heliosphere magnetic field polarity 
of $A>0$, positrons mainly drift from high latitudes to low latitudes, 
while electrons are remain confined to low latitudes and drift along 
the heliospheric current sheet \citep[HCS;][]{2011ApJ...735...83S,
2012Ap&SS.339..223S}. The strength of the heliospheric magnetic field
is weaker in the polar region, and thus positrons are less confined 
by the magnetic field and can reach the Earth more easily. As a result,
the modulation potential $\phi_+$ is smaller than $\phi_-$. For the 
period from 2011 to 2014, the solar activity was at maximum, and the 
effect of drift was not important. Therefore electrons and positrons 
were modulated similarly \citep{2019PhRvD.100d3007V}, with positrons 
being modulated a little bit more than electrons.}

%%%%%%%%%%%%%%%%%%%%%%%%%%%%%%%%%%%%%%%%%%%%%%%%%%%%%%%%%%%%%%
\section{Conclusion and discussion}

With the recent precise measurements of electrons and positions from 
Voyager-1 at outside of the heliosphere and AMS-02 at the TOA, we study
the LIS and the solar modulation effects of electrons and positions. 
We adopt a a non-parametric spline interpolation method to describe
the LIS of ${e^-}$ and ${e^+}$, to minimize the effect of improper
function form assumed. Since there are no measurements of the separate
${e^-}$ and ${e^+}$ fluxes in the Voyager-1 energy window, our resulting
LIS show relatively large uncertainties at low energies. Such LIS may
be used further in the study of electron and positron propagation in
the Milky Way \citep{2015PhRvD..91f3508L}.

We then study the time variations of the ${e^-}$ and ${e^+}$ fluxes 
and the modulation potentials at different time. {\bf We extend the 
simple force-filed approximation to fit the AMS-02 monthly spectra of 
$e^-$ and $e^+$, by multiplying two renormalization factors $c_\pm$.} 
Such renormalization factors show correlations with the solar activity,
and are larger than 1 around the solar maximum and smaller than 1 otherwise.
The renormalization factors might be an empirical approach of a more 
comprehensive solar modulation model other than the force-field
approximation, such as the power-law diffusion model in 
\citet{2019PhRvD.100f3006W} with $K\propto R^{\delta}$, where $K$
is the diffusion coefficient and $R$ is particle rigidity, 
{\bf or even a broken power-law diffusion coefficient 
\citep{2019PhRvD.100d3007V}, or the more complicated model including 
the drift effect \citep{Kuhlen_2019}.}
We leave the detailed investigation in future works.
 
The time variations of fluxes give a time series of the modulation
parameters. Our results show that the modulation potentials for positrons
are well consistent with that for nuclei, which are all positive charge
particles. Nevertheless, the modulation potentials for electrons are
different from those for positrons and nuclei, indicating clearly the
charge-sign dependent modulation. Including the drift effect may 
explain such differences. Although the solar modulation model assumed 
in this work may be over-simplified, we expect that the more complicated 
and physical interpretation of the solar modulation is actually contained 
in the time-variations of the modulation potentials and the renormalization
factors. 

%%%%%%%%%%%%%%%%%%%%%%%%%%%%%%%%%%%%%%%%%%%%%%%%%%%%%%%%%%%%%%

\section*{Acknowledgements}
This works is supported by the National Key Research and Development Program 
of China (No. 2016YFA0400204), the Key Research Program of Frontier Sciences 
of Chinese Academy of Sciences (No. QYZDJ-SSW-SYS024), the National Natural 
Science Foundation of China (Nos. 11851305, 11722328, U1738205).
QY is also supported by the 100 Talents program of Chinese Academy of
Sciences and the Program for Innovative Talents and Entrepreneur in Jiangsu.

\def\aap{A\&A}
\def\apj{ApJ}%
\def\apjl{ApJ}%
\def\apjs{ApJS}%
\def\nar{New A Rev.}%
\def\nat{Nature}%
\def\prl{Phys. Rev. Lett.}
\def\prd{Phys. Rev. D}
\def\apss{Astrophys. Space Sci}
\def\ssr{SSR}

\bibliography{refs}

\appendix

\begin{table}[!htbp]
    \caption{The de-modulated AMS-02 (2019) electron data.}
    \centering
    \renewcommand\arraystretch{0.6}
    \begin{tabular}{ccc}
        \hline\hline
        $E$ & Flux & $\sigma$ \\
        (GeV) & (m$^{-2}$s$^{-1}$sr$^{-1}$GeV$^{-1}$) & (m$^{-2}$s$^{-1}$sr$^{-1}$GeV$^{-1}$) \\
        \hline
        1.362e+00 & 9.884e+01 & 6.048e+00\\ 
        1.522e+00 & 7.434e+01 & 3.808e+00\\ 
        1.702e+00 & 5.801e+01 & 2.516e+00\\ 
        1.902e+00 & 4.587e+01 & 1.715e+00\\ 
        2.122e+00 & 3.541e+01 & 1.149e+00\\ 
        2.372e+00 & 2.705e+01 & 7.625e-01\\ 
        2.642e+00 & 2.095e+01 & 5.260e-01\\ 
        2.942e+00 & 1.574e+01 & 3.546e-01\\ 
        3.262e+00 & 1.203e+01 & 2.462e-01\\ 
        3.612e+00 & 9.030e+00 & 1.710e-01\\ 
        3.962e+00 & 6.882e+00 & 1.221e-01\\ 
        4.332e+00 & 5.324e+00 & 8.943e-02\\ 
        4.712e+00 & 4.131e+00 & 6.618e-02\\ 
        5.112e+00 & 3.191e+00 & 4.864e-02\\ 
        5.552e+00 & 2.449e+00 & 3.624e-02\\ 
        6.032e+00 & 1.867e+00 & 2.701e-02\\ 
        6.532e+00 & 1.430e+00 & 2.060e-02\\ 
        7.052e+00 & 1.106e+00 & 1.528e-02\\ 
        7.602e+00 & 8.580e-01 & 1.161e-02\\ 
        8.182e+00 & 6.668e-01 & 8.916e-03\\ 
        8.782e+00 & 5.206e-01 & 6.966e-03\\ 
        9.412e+00 & 4.071e-01 & 5.330e-03\\ 
        1.007e+01 & 3.236e-01 & 4.271e-03\\ 
        1.075e+01 & 2.569e-01 & 3.311e-03\\ 
        1.146e+01 & 2.053e-01 & 2.738e-03\\ 
        1.220e+01 & 1.645e-01 & 2.195e-03\\ 
        1.298e+01 & 1.327e-01 & 1.696e-03\\ 
        1.378e+01 & 1.076e-01 & 1.405e-03\\ 
        1.461e+01 & 8.775e-02 & 1.153e-03\\ 
        1.548e+01 & 7.152e-02 & 9.430e-04\\ 
        1.638e+01 & 5.870e-02 & 7.742e-04\\ 
        1.731e+01 & 4.812e-02 & 6.407e-04\\ 
        1.827e+01 & 3.998e-02 & 5.342e-04\\ 
        1.927e+01 & 3.325e-02 & 4.408e-04\\ 
        2.030e+01 & 2.786e-02 & 3.706e-04\\ 
        2.137e+01 & 2.314e-02 & 3.143e-04\\ 
        2.247e+01 & 1.938e-02 & 2.675e-04\\ 
        2.362e+01 & 1.632e-02 & 2.216e-04\\ 
        2.480e+01 & 1.378e-02 & 1.887e-04\\ 
        2.604e+01 & 1.169e-02 & 1.669e-04\\ 
        2.735e+01 & 9.795e-03 & 1.359e-04\\ 
        2.874e+01 & 8.292e-03 & 1.160e-04\\ 
        3.022e+01 & 6.983e-03 & 9.828e-05\\ 
        3.179e+01 & 5.816e-03 & 8.179e-05\\ 
        3.345e+01 & 4.915e-03 & 6.980e-05\\ 
        3.522e+01 & 4.131e-03 & 5.859e-05\\ 
        3.711e+01 & 3.457e-03 & 4.978e-05\\ 
        3.912e+01 & 2.923e-03 & 4.238e-05\\ 
        4.127e+01 & 2.436e-03 & 3.567e-05\\ 
        4.357e+01 & 2.024e-03 & 3.034e-05\\ 
        4.605e+01 & 1.693e-03 & 2.510e-05\\ 
        4.871e+01 & 1.413e-03 & 2.136e-05\\ 
        5.159e+01 & 1.154e-03 & 1.711e-05\\ 
        5.471e+01 & 9.572e-04 & 1.476e-05\\ 
        5.811e+01 & 7.876e-04 & 1.235e-05\\ 
        6.182e+01 & 6.353e-04 & 1.012e-05\\ 
        6.590e+01 & 5.196e-04 & 8.374e-06\\ 
        7.041e+01 & 4.168e-04 & 6.763e-06\\ 
        7.544e+01 & 3.366e-04 & 5.614e-06\\ 
        8.108e+01 & 2.665e-04 & 4.473e-06\\ 
        8.748e+01 & 2.086e-04 & 3.567e-06\\ 
        9.481e+01 & 1.606e-04 & 2.829e-06\\ 
        1.034e+02 & 1.233e-04 & 2.237e-06\\ 
        1.135e+02 & 9.016e-05 & 1.688e-06\\ 
        1.258e+02 & 6.542e-05 & 1.261e-06\\ 
        1.409e+02 & 4.582e-05 & 9.370e-07\\ 
        1.597e+02 & 3.040e-05 & 6.509e-07\\ 
        1.839e+02 & 1.869e-05 & 4.380e-07\\ 
        2.170e+02 & 1.083e-05 & 2.881e-07\\ 
        2.626e+02 & 6.014e-06 & 1.805e-07\\ 
        3.276e+02 & 3.144e-06 & 1.111e-07\\ 
        4.293e+02 & 1.283e-06 & 5.657e-08\\ 
        5.896e+02 & 4.572e-07 & 2.940e-08\\ 
        8.331e+02 & 1.774e-07 & 1.704e-08\\ 
        1.179e+03 & 4.129e-08 & 7.348e-09\\

        \hline
    \end{tabular}
    \label{tab:e}
    \end{table}

    \begin{table}[!htbp]
        \caption{The de-modulated AMS-02 (2019) positron data.}
        \renewcommand\arraystretch{0.6}
        \centering
        \begin{tabular}{ccc}
            \hline\hline
            $E$  & {Flux }& $\sigma$ \\
            (GeV) & (m$^{-2}$s$^{-1}$sr$^{-1}$GeV$^{-1}$) & (m$^{-2}$s$^{-1}$sr$^{-1}$GeV$^{-1}$) \\
            \hline 
            1.263e+00 & 1.311e+01 & 1.045e+00\\ 
            1.423e+00 & 9.791e+00 & 5.609e-01\\ 
            1.603e+00 & 7.052e+00 & 3.311e-01\\ 
            1.803e+00 & 5.091e+00 & 2.042e-01\\ 
            2.023e+00 & 3.536e+00 & 1.225e-01\\ 
            2.273e+00 & 2.476e+00 & 7.476e-02\\ 
            2.543e+00 & 1.746e+00 & 4.641e-02\\ 
            2.843e+00 & 1.225e+00 & 2.920e-02\\ 
            3.163e+00 & 8.760e-01 & 1.907e-02\\ 
            3.513e+00 & 6.182e-01 & 1.248e-02\\ 
            3.863e+00 & 4.491e-01 & 8.469e-03\\ 
            4.233e+00 & 3.343e-01 & 6.019e-03\\ 
            4.613e+00 & 2.505e-01 & 4.252e-03\\ 
            5.013e+00 & 1.871e-01 & 3.051e-03\\ 
            5.453e+00 & 1.403e-01 & 2.278e-03\\ 
            5.933e+00 & 1.052e-01 & 1.649e-03\\ 
            6.433e+00 & 7.886e-02 & 1.210e-03\\ 
            6.953e+00 & 6.074e-02 & 9.165e-04\\ 
            7.503e+00 & 4.722e-02 & 7.086e-04\\ 
            8.083e+00 & 3.611e-02 & 5.333e-04\\ 
            8.683e+00 & 2.842e-02 & 4.220e-04\\ 
            9.313e+00 & 2.218e-02 & 3.306e-04\\ 
            9.973e+00 & 1.767e-02 & 2.573e-04\\ 
            1.065e+01 & 1.421e-02 & 2.128e-04\\ 
            1.136e+01 & 1.142e-02 & 1.716e-04\\ 
            1.210e+01 & 9.341e-03 & 1.428e-04\\ 
            1.288e+01 & 7.727e-03 & 1.186e-04\\ 
            1.368e+01 & 6.289e-03 & 9.868e-05\\ 
            1.451e+01 & 5.120e-03 & 8.104e-05\\ 
            1.538e+01 & 4.237e-03 & 6.774e-05\\ 
            1.628e+01 & 3.559e-03 & 5.815e-05\\ 
            1.721e+01 & 2.951e-03 & 4.939e-05\\ 
            1.817e+01 & 2.478e-03 & 4.176e-05\\ 
            1.917e+01 & 2.081e-03 & 3.517e-05\\ 
            2.020e+01 & 1.786e-03 & 3.105e-05\\ 
            2.127e+01 & 1.554e-03 & 2.701e-05\\ 
            2.237e+01 & 1.293e-03 & 2.367e-05\\ 
            2.352e+01 & 1.081e-03 & 1.974e-05\\ 
            2.470e+01 & 9.560e-04 & 1.798e-05\\ 
            2.594e+01 & 8.073e-04 & 1.549e-05\\ 
            2.725e+01 & 7.114e-04 & 1.386e-05\\ 
            2.864e+01 & 6.036e-04 & 1.209e-05\\ 
            3.012e+01 & 5.304e-04 & 1.071e-05\\ 
            3.169e+01 & 4.466e-04 & 9.349e-06\\ 
            3.335e+01 & 3.839e-04 & 8.295e-06\\ 
            3.512e+01 & 3.253e-04 & 7.175e-06\\ 
            3.701e+01 & 2.860e-04 & 6.503e-06\\ 
            3.902e+01 & 2.413e-04 & 5.452e-06\\ 
            4.117e+01 & 2.073e-04 & 5.016e-06\\ 
            4.347e+01 & 1.779e-04 & 4.426e-06\\ 
            4.595e+01 & 1.491e-04 & 3.839e-06\\ 
            4.861e+01 & 1.362e-04 & 3.545e-06\\ 
            5.149e+01 & 1.057e-04 & 2.965e-06\\ 
            5.461e+01 & 9.089e-05 & 2.608e-06\\ 
            5.801e+01 & 7.742e-05 & 2.283e-06\\ 
            6.172e+01 & 6.255e-05 & 1.949e-06\\ 
            6.580e+01 & 5.620e-05 & 1.768e-06\\ 
            7.031e+01 & 4.454e-05 & 1.475e-06\\ 
            7.534e+01 & 3.897e-05 & 1.295e-06\\ 
            8.098e+01 & 3.065e-05 & 1.084e-06\\ 
            8.738e+01 & 2.551e-05 & 9.272e-07\\ 
            9.471e+01 & 2.067e-05 & 7.762e-07\\ 
            1.033e+02 & 1.481e-05 & 6.495e-07\\ 
            1.134e+02 & 1.187e-05 & 5.264e-07\\ 
            1.257e+02 & 8.774e-06 & 4.097e-07\\ 
            1.408e+02 & 7.067e-06 & 3.311e-07\\ 
            1.596e+02 & 4.635e-06 & 2.385e-07\\ 
            1.838e+02 & 3.225e-06 & 1.750e-07\\ 
            2.169e+02 & 1.883e-06 & 1.185e-07\\ 
            2.625e+02 & 1.164e-06 & 8.170e-08\\ 
            3.275e+02 & 5.798e-07 & 5.204e-08\\ 
            4.292e+02 & 2.499e-07 & 3.011e-08\\ 
            5.895e+02 & 8.332e-08 & 1.848e-08\\ 
            8.330e+02 & 1.930e-08 & 1.176e-08\\

            \hline
        \end{tabular}
        \label{tab:p}
        \end{table} 

\end{document}